\documentclass[letter]{aa}  

\usepackage[varg]{txfonts}

\usepackage{siunitx}
\usepackage{lipsum}
\usepackage{here}

\usepackage[colorlinks=true, citecolor=blue]{hyperref}
\begin{document}

   \title{Threshold velocity for collisional growth of porous dust aggregates consisting of cohesive frictionless spheres}

   \titlerunning{Collisional growth of porous dust aggregates consisting of cohesive frictionless spheres}

   \author{Sota Arakawa
          \inst{1}
          \and
          Hidekazu Tanaka
          \inst{2}
          \and
          Eiichiro Kokubo
          \inst{3}
          \and
          Daisuke Nishiura
          \inst{1}
          \and
          Mikito Furuichi
          \inst{1}
          }

   \institute{Japan Agency for Marine-Earth Science and Technology, 3173-25, Showa-machi, Kanazawa-ku, Yokohama, 236-0001, Japan\\
              \email{arakawas@jamstec.go.jp}
         \and
             Astronomical Institute, Graduate School of Science, Tohoku University, 6-3 Aramaki, Aoba-ku, Sendai, 980-8578, Japan
         \and
             National Astronomical Observatory of Japan, 2-21-1, Osawa, Mitaka, Tokyo, 181-8588, Japan
             }

   \date{Received DD/MM/YYYY; accepted DD/MM/YYYY}

 
  \abstract{
  Understanding the collisional outcomes of dust aggregates and dependence on material properties of the constituting particles is of great importance toward understanding planet formation.
  Recent numerical simulations have revealed that interparticle tangential friction plays a crucial role in energy dissipation during collisions between porous dust aggregates; however, the importance of friction on the collisional growth of dust aggregates remains poorly understood.
  Here we demonstrate the effects of interparticle tangential friction on the collisional growth of dust aggregates.
  We performed numerical simulations of collisions between equal-mass porous dust aggregates consisting of cohesive and frictionless spheres.
  We changed the collision velocity and impact angle systematically and calculated the collisional growth efficiency as a function of the collision velocity.
  We found that the threshold velocity for collisional growth decreases when dust aggregates are made of frictionless spheres as compared to frictional spheres.
  Our results highlight the importance of tangential interactions on the collisional behavior of dust aggregates and indicate that the predictive equation for threshold velocity should be reconstructed.
  }


   \maketitle
%

\section{Introduction}

Collisional growth of dust aggregates consisting of micro/nanosized grains is ubiquitous in the universe.
For example, collisional growth of soot nanoparticles produced by combustion processes is important in environmental science and engineering \citep[e.g.,][]{haynes1981soot}.
Atmospheric hazes on the early Earth and other planetary bodies are also dust aggregates consisting of micro/nanosized particles, and their collisional growth affects the atmospheric structure \citep[e.g.,][]{2006PNAS..10318035T, 2017Natur.551..352Z, 2021ApJ...912...37O}.
The first step of planet formation in circumstellar disks is collisional growth of micro/nanosized interstellar dust particles \citep[e.g.,][]{1985prpl.conf.1100H, 2022arXiv220309759D}, and dust growth might also trigger the formation of planetary objects around supermassive black holes in a galactic center \citep[e.g.,][]{2021ApJ...909...96W}.

The condition for collisional growth of dust aggregates has been studied both by laboratory experiments \citep[e.g.,][]{2008ARA&A..46...21B, 2010A&A...513A..56G, 2021ApJ...923..134F, 2022MNRAS.509.5641S} and numerical simulations \citep[e.g.,][]{2009ApJ...702.1490W, 2013A&A...560A..45S, 2021ApJ...915...22H, OSINSKY2022127785}.
These previous studies found that the growth/fragmentation conditions depend on the strength of interparticle forces acting on constituent particles in contact with each other.

\citet{2009ApJ...702.1490W} performed a large number of numerical simulations of collisions between two equal-mass dust aggregates consisting of submicron-sized spherical ice particles.
They varied the collision velocities and impact angles and obtained the collisional growth efficiency.
Their numerical simulations revealed that the threshold velocity for collisional growth/fragmentation of dust aggregates, $v_{\rm fra}$, is approximately $60~\si{m}~\si{s}^{-1}$ when the radius of constituting particles is $0.1~\si{\micro m}$ and the initial aggregates before collisions are prepared by ballistic particle--cluster aggregation \citep[BPCA;][]{1992A&A...262..315M}.
They also predicted that $v_{\rm fra}$ would be proportional to the square root of the energy required for breaking one interparticle contact, $E_{\rm break}$, which is related to the connection/disconnection of particles due to interparticle normal motion \citep[e.g.,][]{1997ApJ...480..647D, 2007ApJ...661..320W}.

\citet{2022ApJ...933..144A} also performed numerical simulations of collisions between dust aggregates and investigated the dependence of the threshold velocity on the strength of the viscous dissipation.
They found that the main energy dissipation mechanism is not the normal interaction (i.e., viscous dissipation and connection/disconnection of particles) but the rolling friction between particles in contact.

Then \citet{2022ApJ...939..100A} investigated the dependence of the threshold velocity on the strength of rolling friction.
They revealed that the threshold velocity barely depends on the strength of rolling friction, however.
In their simulations, frictions arising from two other tangential motions (sliding and twisting) complemented the change in energy dissipation due to rolling friction.

Energy dissipation is essentially required for collisional growth of dust aggregates.
The results of \citet{2022ApJ...933..144A, 2022ApJ...939..100A} indicate that the threshold velocity for collisional growth/fragmentation are affected by the presence/absence of interparticle tangential frictions when the motions cannot complement one another.

Here, we report numerical simulations of collisions between two equal-mass dust aggregates consisting of submicron-sized spherical ice particles, with and without interparticle frictions.
We also systematically varied the collision velocities and impact angles, and we calculated the collisional growth efficiency by averaging over impact angles.
We confirmed that the collisional growth efficiency of dust aggregates clearly depends on the presence/absence of interparticle frictions associated with tangential motions.
Our results indicate that the dependence of the collisional growth efficiency on the material properties of constituent particles are more complex than previously assumed.

\section{Model}

We performed three-dimensional numerical simulations of collisions between two equal-mass dust aggregates.
Our numerical code was originally developed by \citet{2007ApJ...661..320W}, and \citet{2022ApJ...933..144A} introduced a viscous drag force for normal motion \citep[e.g.,][]{2013JPhD...46Q5303K}.
We prepared two initial aggregates before collisions by BPCA as in previous studies \citep{2009ApJ...702.1490W, 2022ApJ...933..144A, 2022ApJ...939..100A}.
We set the number of particles in the target aggregate, $N_{\rm tar}$, equal to that for the projectile aggregate, $N_{\rm pro}$, and the total number of particles in a simulation, $N_{\rm tot} = N_{\rm tar} + N_{\rm pro}$, is 100,000.
The constituent dust particles are made of water ice and the particle radius is $r_{1} = 0.1~\si{\micro m}$.
The particle interaction model is briefly described in Appendix \ref{app:model}.

The strength of interparticle normal dissipation is controlled by a parameter called the viscoelastic timescale, $T_{\rm vis}$ \citep[e.g.,][]{2022ApJ...933..144A}. 
We performed numerical simulations with two models for interparticle normal dissipation: with dissipation ($T_{\rm vis} = 6~\si{ps}$) and without dissipation ($T_{\rm vis} = 0~\si{ps}$; see Appendix \ref{app:model-n}).

The strengths of interparticle tangential frictions are characterized by the spring constants for rolling ($k_{\rm r}$), sliding ($k_{\rm s}$), and twisting ($k_{\rm t}$).
In this study, we also performed numerical simulations with two models for interparticle tangential motions: the frictional and frictionless models.
In the frictional model, we consider the interparticle tangential interactions as modeled by \citet{2007ApJ...661..320W}, and we used the same values of $k_{\rm r}$, $k_{\rm s}$, and $k_{\rm t}$ as assumed in \citet{2007ApJ...661..320W}.
In the frictionless model, in contrast, we do not consider the interparticle tangential frictions; in other words, we set $k_{\rm r} = k_{\rm s} = k_{\rm t} = 0$ (see Appendix \ref{app:model-t}).

\section{Results}

Figure \ref{fig:snapshot} shows snapshots of the collisional outcomes.
We show the numerical results for the frictionless model ($k_{\rm r} = k_{\rm s} = k_{\rm t} = 0$) without normal dissipation ($T_{\rm vis} = 0~\si{ps}$), and we set $v_{\rm col} = 39.8~\si{m.s^{-1}}$, where $v_{\rm col}$ is the collision velocity of two dust aggregates.
We found that the collisional behavior strongly depends on the impact angles.
Comparing Figure \ref{fig:snapshot}(c) with Figure 2 of \citet{2022ApJ...939..100A}, we can visually understand that the collisional behavior also depends on the interparticle interaction model.
\citet{2022ApJ...939..100A} presented the results for the same collision velocity and impact angle, but used the frictional model with normal dissipation.
In contrast, Figure \ref{fig:snapshot}(c) of this study shows the results for the frictionless model without normal dissipation.
We found that the amount of small fragments significantly increases as compared to the results of \citet{2022ApJ...939..100A}.

\begin{figure}
\centering
\includegraphics[width=\columnwidth]{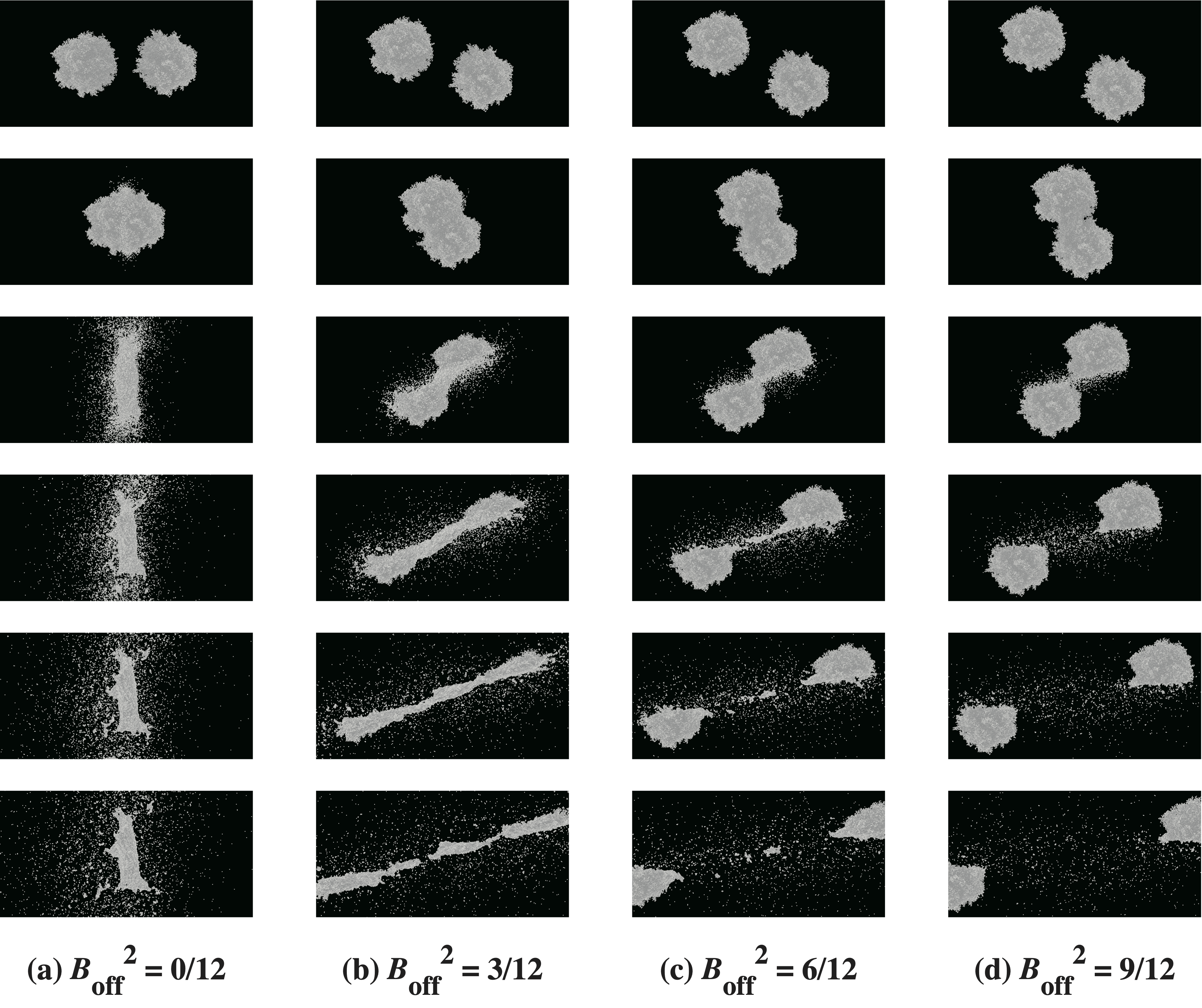}
\caption{
Snapshots of the collisional outcomes.
Here, we show the numerical results for the frictionless model ($k_{\rm r} = k_{\rm s} = k_{\rm t} = 0$) without normal dissipation ($T_{\rm vis} = 0~\si{ps}$), and we set $v_{\rm col} = 39.8~\si{m.s^{-1}}$.
Panels (a)--(d) are the time series of snapshots for ${B_{\rm off}}^{2} = 0 / 12$, $3 / 12$, $6 / 12$, and $9 / 12$, respectively.
The time interval for each snapshot is $0.40~\si{\micro s}$.
}
\label{fig:snapshot}
\end{figure}

In this study, we use the normalized impact parameter, $B_{\rm off}$, to quantify the offset of oblique collisions. 
We defined $B_{\rm off}$ as $B_{\rm off} = b_{\rm off} / b_{\rm max}$, where $b_{\rm off}$ is the impact parameter and $b_{\rm max}$ is the sum of the radii of the target and projectile aggregates \citep[see][]{2022ApJ...933..144A}.
The radius of dust aggregates is set equal to the characteristic radius, $r_{\rm c}$, which is given by $r_{\rm c} = \sqrt{5/3} r_{\rm g}$, where $r_{\rm g}$ is the gyration radius \citep[e.g.,][]{1992A&A...262..315M, 2009ApJ...702.1490W}.
It is clear that ${B_{\rm off}}^{2} = 0$ for head-on collisions, and ${B_{\rm off}}^{2}$ ranges from 0 to 1.

When considering collisions in space, the impact parameter should vary with each collision event.
The average value of a variable $A$ weighted over $B_{\rm off}$,
\begin{equation}
{\langle A \rangle} \equiv \int_{0}^{1} {\rm d}{B_{\rm off}}~2 B_{\rm off} A,
\end{equation}
is useful to measure the average outcome of the collision.

\subsection{Collisional growth efficiency and the threshold velocity for collisional growth/fragmentation}

\begin{figure}
\centering
\includegraphics[width=\columnwidth]{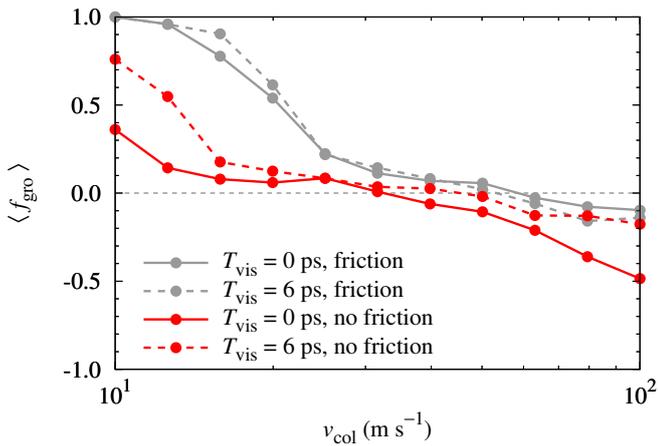}
\caption{
$B_{\rm off}$-weighted collisional growth efficiency, ${\langle f_{\rm gro} \rangle}$.
We note that the results for frictional models (gray lines) are identical to those presented in Figure 3 of \citet{2022ApJ...933..144A} for $v_{\rm col} \ge 20.0~\si{m}~\si{s}^{-1}$.
}
\label{fig:1st_ave}
\end{figure}

Figure \ref{fig:1st_ave} shows the $B_{\rm off}$-weighted collisional growth efficiency, ${\langle f_{\rm gro} \rangle}$, as a function of $v_{\rm col}$ for different particle interaction models.
The growth efficiency, $f_{\rm gro}$, is defined for each simulation, and it is given by $f_{\rm gro} = {( N_{\rm lar} - N_{\rm tar} )} / N_{\rm pro}$, where $N_{\rm lar}$ denotes the number of constituent particles in the largest remnant \citep[see][]{2021ApJ...915...22H}.
The collisional outcome of each simulation is summarized in Figure \ref{fig:1st} (see Appendix \ref{app:1st}).

We found that ${\langle f_{\rm gro} \rangle}$ clearly depends on the choice of particle interaction models.
In our simulation of the frictional model without normal dissipation (gray solid line), ${\langle f_{\rm gro} \rangle} \simeq 0$ at approximately $v_{\rm col} = 60~\si{m}~\si{s}^{-1}$.
\citet{2009ApJ...702.1490W} also performed simulations with the same particle interaction model and their result is in excellent agreement with ours.
Here we define the threshold velocity for collisional growth/fragmentation, $v_{\rm fra}$, as the collision velocity that satisfies ${\langle f_{\rm gro} \rangle} = 0$.

In the frictional model with normal dissipation (gray dashed line), we found that $v_{\rm fra} \simeq 50~\si{m}~\si{s}^{-1}$, which is close to the value for the frictional model without normal dissipation.
The dependence of ${\langle f_{\rm gro} \rangle}$ on $v_{\rm col}$ in frictional models with and without normal dissipation is similar; their difference in ${\langle f_{\rm gro} \rangle}$ is typically less than $0.1$ in the range of $10~\si{m}~\si{s}^{-1} \le v_{\rm col} \le 100~\si{m}~\si{s}^{-1}$.
\citet{2022ApJ...933..144A} concluded that this independence of ${\langle f_{\rm gro} \rangle}$ on the presence/absence of normal dissipation is consistent with the fact that the main energy dissipation mechanism is not normal dissipation but interparticle tangential friction when using frictional models for particle interaction.

In frictionless models, however, we found that $v_{\rm fra}$ is significantly lower than that of frictional models, e.g., $v_{\rm fra} \simeq 30~\si{m}~\si{s}^{-1}$ in the frictionless model without normal dissipation (red solid line).
In addition, ${\langle f_{\rm gro} \rangle}$ also decreases compared to the frictional models.
For example, $0.1 \lesssim {\langle f_{\rm gro} \rangle} \lesssim 0.2$ at $v_{\rm col} = 15.8~\si{m}~\si{s}^{-1}$ in frictionless models, while $0.8 \lesssim {\langle f_{\rm gro} \rangle} \lesssim 0.9$ at the same $v_{\rm col}$ in frictional models.
These differences would be related to the difference in energy dissipation processes (see Appendix \ref{app:edis}).

We note that $v_{\rm fra}$ depends on the presence/absence of normal dissipation when tangential frictions are absent.
In the frictionless model with normal dissipation (red dashed line), our numerical results show that $v_{\rm fra} \simeq 45~\si{m}~\si{s}^{-1}$, which is 1.5 times larger than that of the frictionless model without normal dissipation.
Figure \ref{fig:1st_ave} shows that ${\langle f_{\rm gro} \rangle}$ for the frictionless model with normal dissipation is significantly higher than that of the frictionless model without normal dissipation in the ranges of $v_{\rm col} \ll 20~\si{m}~\si{s}^{-1}$ and $v_{\rm col} \gg 60~\si{m}~\si{s}^{-1}$, although their difference in ${\langle f_{\rm gro} \rangle}$ is roughly within $0.1$ in the intermediate range of $v_{\rm col}$.

Based on the results of \citet{2009ApJ...702.1490W}, \citet{2013A&A...559A..62W} proposed an empirical formula to estimate $v_{\rm fra}$ as a function of the particle radius and material properties of constituting particles.
The empirical formula is $v_{\rm fra} \simeq 15 \sqrt{E_{\rm break} / m_{1}}$, where $m_{1}$ is the mass of each particle.
As both $E_{\rm break}$ and $m_{1}$ are independent of the spring constants for tangential motions ($k_{\rm r}$, $k_{\rm s}$, and $k_{\rm t}$), this equation cannot express the effects of tangential interactions on $v_{\rm fra}$.
Our numerical results, however, highlight the impact of tangential interactions on $v_{\rm fra}$.
Thus we need to modify the prediction formula for $v_{\rm fra}$.

Our results indicate that $v_{\rm fra}$ depends not only on $E_{\rm break}$ but also on interparticle energies associated with tangential motions.
It is important to note that their dependences on the particle radius and material properties are different from each other \citep[see][]{2007ApJ...661..320W}.
For example, $E_{\rm break}$ is proportional to ${r_{1}}^{4/3}$ (see Appendix \ref{app:model-n}), while the energy needed to slide a particle by $\pi / 2$ radian around its contact point, $E_{\rm slide}$, is proportional to ${r_{1}}^{7/3}$.
The energy needed to twist over $\pi / 2$ radian, $E_{\rm twist}$, is proportional to ${r_{1}}^{2}$ (see Appendix \ref{app:model-t}).
Although it seems an extreme and unrealistic assumption, $v_{\rm fra}$ might be proportional to ${r_{1}}^{- 1/3}$ when it is proportional to $\sqrt{E_{\rm slide} / m_{1}}$, or $v_{\rm fra}$ might be proportional to ${r_{1}}^{- 1/2}$ when it is proportional to $\sqrt{E_{\rm twist} / m_{1}}$.

\citet{2008ARA&A..46...21B} reviewed laboratory experiments of collisions of dust aggregates and reported that the threshold velocity for sticking of dust aggregates is proportional to ${r_{1}}^{-x}$, with $x \gtrsim 1$ for $0.1~\si{\micro m} \lesssim r_{1} \lesssim 1~\si{\micro m}$ and $x \lesssim 1$ for $1~\si{\micro m} \lesssim r_{1} \lesssim 10~\si{\micro m}$.
Although these experimental results are for head-on collisions and it cannot be directly compared with our $B_{\rm off}$-weighted numerical results, we can speculate that the dependence of $v_{\rm fra}$ on $r_{1}$ might not be given by a simple power-law relation.

\subsection{Size distribution of fragments}

As shown in Figure \ref{fig:snapshot}, collisions of dust aggregates consisting of frictionless particles without normal dissipation ($k_{\rm r} = k_{\rm s} = k_{\rm t} = 0$ and $T_{\rm vis} = 0~\si{ps}$) produce a large amount of small fragments.
In circumstellr disks, the strength of turbulence driven by magnetorotational instability is a function of the gas ionization degree and depends on the amount of small dust aggregates \citep[e.g.,][]{2012ApJ...753L...8O}.
Thus, the amount of small dust aggregates might also be the key parameter for planet formation via collisional growth of dust aggregates.
Here, we quantify the size distribution of fragments in our simulations \citep[see also][]{2022ApJ...933..144A, 2022arXiv221210796H, OSINSKY2022127785}.

\begin{figure}
\centering
\includegraphics[width=\columnwidth]{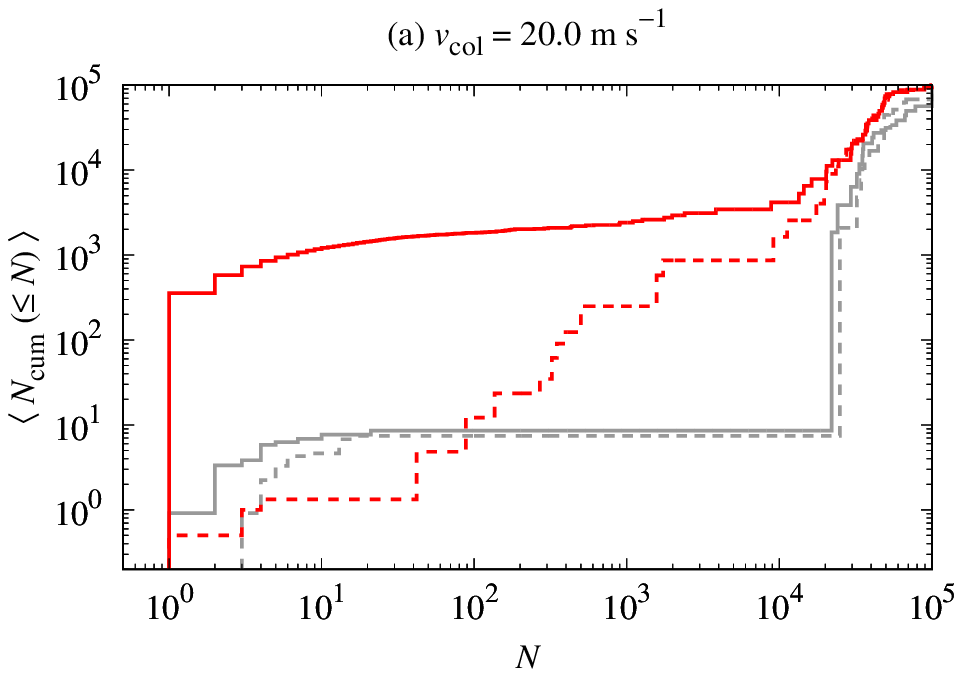}
\includegraphics[width=\columnwidth]{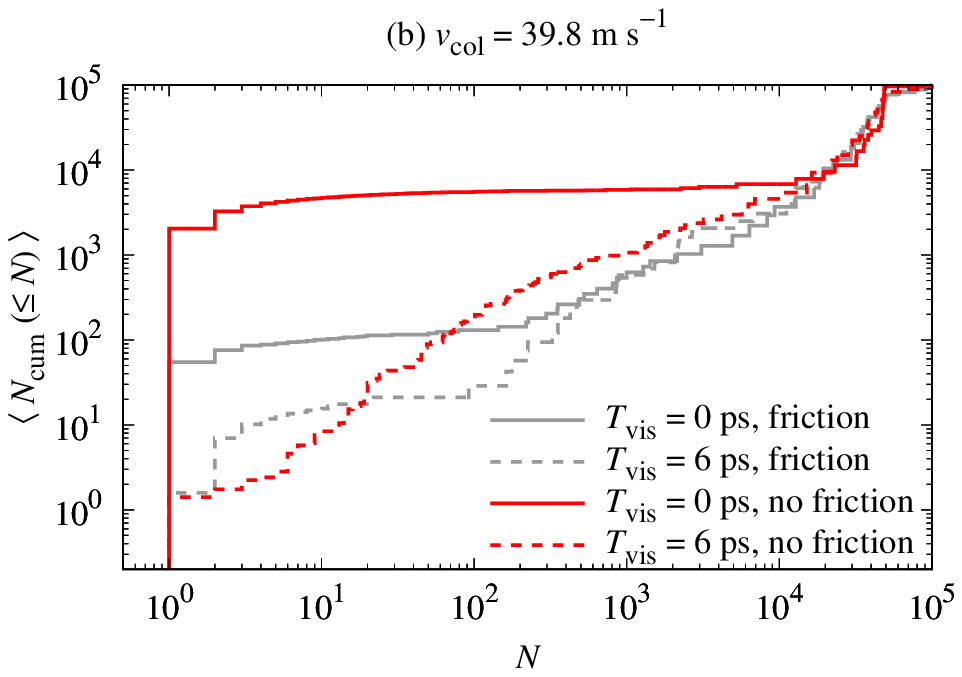}
\caption{
$B_{\rm off}$-weighted average of ${N_{\rm cum} (\le N)}$, ${\langle N_{\rm cum} (\le N) \rangle}$.
(a) For the case of $v_{\rm col} = 20.0~\si{m}~\si{s}^{-1}$.
(c) For the case of $v_{\rm col} = 39.8~\si{m}~\si{s}^{-1}$.
}
\label{fig:m_cum_ave}
\end{figure}

\citet{2022ApJ...933..144A} defined ${N_{\rm cum} (\le N)}$ as the cumulative number of particles that are constituents of fragments that contain not larger than $N$ particles.
Figure \ref{fig:m_cum_ave} shows the $B_{\rm off}$-weighted average of ${N_{\rm cum} (\le N)}$, ${\langle N_{\rm cum} (\le N) \rangle}$.
We found that ${\langle N_{\rm cum} (\le N) \rangle}$ significantly differs among particle interaction models.

As shown in Figure \ref{fig:m_cum_ave}, the size distributions of fragments for frictional models with and without normal dissipations (gray dashed and solid lines, respectively) are similar, particularly for larger fragments ($N > 10$ for $v_{\rm col} = 20.0~\si{m}~\si{s}^{-1}$ and $N > 10^{3}$ for $v_{\rm col} = 39.8~\si{m}~\si{s}^{-1}$). 
This trend is consistent with the results of \citet{2022ApJ...933..144A}.

We found that the amount of small fragments strongly depends on the presence/absence of interparticle tangential friction.
When we apply the frictionless model without normal dissipation for constituting particles (red solid line), the amount of small fragments with $N < 10^{2}$ was orders of magnitude larger than that of the frictional models.
This difference might also imply the importance of interparticle tangential frictions on the energy dissipation and deformation of dust aggregates during collision.

We note that for frictionless models, the size distribution of fragments clearly depends on the strength of normal dissipation ($T_{\rm vis}$).
Considering $v_{\rm col} = 20.0~\si{m}~\si{s}^{-1}$ as an example, the difference in ${\langle f_{\rm gro} \rangle}$ is small (approximately $0.1$; Figure \ref{fig:1st_ave}), but the difference in ${\langle N_{\rm cum} (\le N) \rangle}$ is approximately two orders of magnitude for $N < 10^{2}$.
These simulation results might indicate that the mechanisms for aggregate-wide deformation and the ejection of small fragments are different.
We will test this hypothesis in future studies.

\section{Conclusions}

Understanding the collisional behavior of dust aggregates consisting of micro/nanosized grains is essential to understanding planet formation.
The threshold velocity for collisional growth/fragmentation, $v_{\rm fra}$, is one of the most important parameters that controls the size of dust aggregates in circumstellar disks \citep[e.g.,][]{2019ApJ...878..132O, 2021ApJ...920...27A}; however, the dependence of $v_{\rm fra}$ on the particle radius and the material properties of constituting particles is still under debate.

\citet{2022ApJ...933..144A, 2022ApJ...939..100A} revealed that the main energy dissipation mechanism for oblique collisions of dust aggregates is not the interparticle normal interaction (i.e., connection and disconnection of particles) but the tangential friction between particles in contact with each other.
Thus, we expect that the collisional outcomes of dust aggregates could depend on the strength of interparticle frictions.

In this study, we demonstrated that $v_{\rm fra}$ depends on the strength of tangential interactions using numerical simulations of collisions between dust aggregates.
We tested $2 \times 2 = 4$ types of particle interaction models in this study.
In these models, the presence/absence of the interparticle tangential frictions is captured by differences in the spring constants ($k_{\rm r}$, $k_{\rm s}$, and $k_{\rm t}$), and the presence/absence of the normal dissipation is captured by differences in the viscoelastic timescale ($T_{\rm vis}$).

In the frictional model without normal dissipation, we found that $v_{\rm fra} \simeq 60~\si{m}~\si{s}^{-1}$, which is consistent with that reported in previous studies \citep[e.g.,][]{2009ApJ...702.1490W, 2021ApJ...915...22H}.
In contrast, in the frictionless model without normal dissipation, we found that $v_{\rm fra} \simeq 30~\si{m}~\si{s}^{-1}$, which is notably lower than that of the frictional model (see Figure \ref{fig:1st_ave}).
We also found that $v_{\rm fra}$ depends on the presence/absence of the normal dissipation when tangential frictions are absent, while $v_{\rm fra}$ is nearly independent of $T_{\rm vis}$ for frictional models.
Our results further indicate that the dependence of $v_{\rm fra}$ on the particle radius and material properties cannot be described by a simple power-law relation \citep[see][]{2008ARA&A..46...21B, 2009ApJ...702.1490W, 2013A&A...559A..62W}.
Future studies on this point are essential, although a large number of numerical simulations are needed to construct a better fitting formula to predict $v_{\rm fra}$ as a function of particle radius and material properties.

The size distribution of fragments also depends on the choice of particle interaction models (see Figure \ref{fig:m_cum_ave}).
As shown in Figure \ref{fig:snapshot}, collisions of dust aggregates consisting of frictionless particles without normal dissipation produce a large amount of small fragments.
We also found that for frictionless models, the size distribution of fragments significantly depends on the strength of normal dissipation, even if the difference in ${\langle f_{\rm gro} \rangle}$ is small.
Our results might indicate that the mechanisms for aggregate-wide deformation and the ejection of small fragments are different, although we need to investigate this hypothesis in future studies.

\begin{acknowledgements}
The anonymous reviewer provided a constructive review that improved this paper.
Numerical computations were carried out on PC cluster at CfCA, NAOJ.
H.T.\ and E.K.\ were supported by JSPS KAKENHI grant No.\ 18H05438.
We thank American Journal Experts (AJE) for English language editing.
\end{acknowledgements}

%
%

\bibliographystyle{aa}
\bibliography{sample631}


\appendix

\section{Particle interaction model}
\label{app:model}

Here, we briefly explain the particle interaction model used in this study.
Details are described in \citet{2007ApJ...661..320W} and \citet{2022ApJ...933..144A}.

\subsection{Normal motion}
\label{app:model-n}

We assume that the normal force acting between two particles, $F$, is given by the sum of the two terms:
\begin{equation}
F = F_{\rm E} + F_{\rm D},
\end{equation}
where $F_{\rm E}$ denotes the force arising from the elastic deformation of particles and $F_{\rm D}$ is the force related to the viscous dissipation.

When two particles in contact are elastic spheres with a surface energy, the elastic term is given by the following equation \citep{1971RSPSA.324..301J}:
\begin{equation}
F_{\rm E} = 4 {\left[ {\left( \frac{a}{a_{0}} \right)}^{3} - {\left( \frac{a}{a_{0}} \right)}^{3/2} \right]} F_{\rm c},
\end{equation}
where $a$ is the contact radius, $a_{0}$ is the contact radius at equilibrium, $F_{\rm c} = 3 \pi \gamma R$ is the maximum force needed to separate the two particles in contact, $\gamma$ is the surface energy, and $R = r_{1} / 2$ is the reduced particle radius.
At the equilibrium state, $a_{0}$ is given by
\begin{equation}
a_{0} = {\left( \frac{9 \pi \gamma R^{2}}{\mathcal{E}^{*}} \right)}^{1/3},
\end{equation}
where $\mathcal{E}^{*}$ is the reduced Young's modulus.

The contact radius is a function of the compression length between two particles in contact, $\delta$, and vice versa:
\begin{equation}
\frac{\delta}{\delta_{0}} = 3 {\left( \frac{a}{a_{0}} \right)}^{2} - 2 {\left( \frac{a}{a_{0}} \right)}^{1/2},
\end{equation}
where $\delta_{0} = {a_{0}}^{2} / {( 3 R )}$ is the equilibrium compression length at $a = a_{0}$.
Two particles in contact separate when the compression length reaches the critical length, $\delta = - \delta_{\rm c}$, where $\delta_{\rm c} = {( 9 / 16 )}^{1/3} \delta_{0}$.

The viscous drag force is given by
\begin{equation}
F_{\rm D} = \frac{2 T_{\rm vis} \mathcal{E}^{*}}{\nu^{2}} a v_{\rm rel},
\end{equation}
where $\nu$ is Poisson's ratio, $v_{\rm rel}$ is the normal component of the relative velocity of the two particles, and $T_{\rm vis}$ is the viscoelastic timescale \citep[see][]{2013JPhD...46Q5303K}.

In this study, we consider two cases: $T_{\rm vis} = 0~\si{ps}$ and $T_{\rm vis} = 6~\si{ps}$.
We set $T_{\rm vis} = 6~\si{ps}$ in our previous studies \citep[e.g.,][]{2022ApJ...939..100A}, and the choice of $T_{\rm vis} = 6~\si{ps}$ is motivated by extrapolation of laboratory experiments \citep[see Figure 5 of][]{2021ApJ...910..130A}.
\citet{2013JPhD...46Q5303K} predicted that $T_{\rm vis}$ would be approximately proportional to $R$, and \citet{2021ApJ...910..130A} confirmed this relation using experimental results of \citet{2015ApJ...798...34G} and \citet{2016ApJ...827...63M}.

The potential energy for normal motion of the two particles in contact, $U_{\rm n}$, is given by
\begin{equation}
\frac{U_{\rm n}}{F_{\rm c} \delta_{\rm c}} = 4 \times 6^{1/3} \times {\left[ \frac{4}{5} {\left( \frac{a}{a_{0}} \right)}^{5} - \frac{4}{3} {\left( \frac{a}{a_{0}} \right)}^{7/2} + \frac{1}{3} {\left( \frac{a}{a_{0}} \right)}^{2} \right]}.
\end{equation}
The energy needed to break a contact in equilibrium by a quasistatic process (i.e., $v_{\rm rel} \to 0$ and $F_{\rm D} \to 0$), $E_{\rm break}$, is
\begin{eqnarray}
E_{\rm break} & = & U_{\rm n} {( - \delta_{\rm c} )} - U_{\rm n} {( \delta_{0} )} \nonumber \\
              & = & {\left( \frac{4}{45} + \frac{4}{5} \times 6^{1/3} \right)} F_{\rm c} \delta_{\rm c},
\end{eqnarray}
and $E_{\rm break}$ is proportional to ${r_{1}}^{4/3}$.

\subsection{Tangential motion}
\label{app:model-t}

The tangential motion of two particles in contact is the combination of three motions: rolling, sliding, and twisting.
The displacements corresponding to these motions are described as the rotation of two particles in contact.
\citet{2007ApJ...661..320W} provided the particle interaction model for these tangential motions \citep[see also][]{1995PMagA..72..783D,1996PMagA..73.1279D}, which is equivalent to the linear spring model with critical displacements to their elastic limits.
The concept of the tangential interaction model is summarized in Figures 2 and 3 of \citet{2007ApJ...661..320W}.

\subsubsection{Rolling motion}

The spring constant for the rolling displacement, $\xi$, is $k_{\rm r}$.
For the frictional model, we set
\begin{equation}
k_{\rm r} = \frac{4 F_{\rm c}}{R},
\end{equation}
and we set $k_{\rm r} = 0$ for the frictionless model.
The energy needed to rotate a particle by $\pi / 2$ radian around its contact point, $E_{\rm roll}$, is useful to interpret the collisional outcomes of dust aggregates from energetics \citep[e.g.,][]{1997ApJ...480..647D}.
\citet{2007ApJ...661..320W} derived that $E_{\rm roll}$ is given by
\begin{eqnarray}
E_{\rm roll} & = & k_{\rm r} \xi_{\rm crit} \pi R \nonumber \\
             & = & 12 \pi^{2} \gamma R \xi_{\rm crit},
\end{eqnarray}
where $\xi_{\rm crit}$ is the critical rolling displacement.
In this study, we set $\xi_{\rm crit} = 0.8~{\rm nm}$ for water ice particles of $r_{1} = 0.1~\si{\micro m}$.
The dependence of $\xi_{\rm crit}$ on $r_{1}$ is poorly understood, however \citep[see][and references therin]{2022ApJ...939..100A}.

\subsubsection{Sliding motion}

The spring constant for the sliding displacement, $\zeta$, is $k_{\rm s}$.
For the frictional model, we set
\begin{equation}
k_{\rm s} = 8 a_{0} \mathcal{G}^{\star},
\end{equation}
where $\mathcal{G}^{\star} = \mathcal{G} / {[ 2 {( 2 - \nu )} ]}$, and $\mathcal{G}$ is the shear modulus.
We set $k_{\rm s} = 0$ for the frictionless model, as is the case for $k_{\rm r}$.
The critical sliding displacement, $\zeta_{\rm crit}$, is given by $\zeta_{\rm crit} = {[ {( 2 - \nu )} / {( 16 \pi )} ]} a_{0}$ \citep{2007ApJ...661..320W}.
The energy needed to slide a particle by $\pi / 2$ radian around its contact point, $E_{\rm slide}$, is given by
\begin{eqnarray}
E_{\rm slide} & = & k_{\rm s} \zeta_{\rm crit} \pi R \nonumber \\
              & = & \frac{1}{4} \mathcal{G} {a_{0}}^{2} R,
\end{eqnarray}
and $E_{\rm slide}$ is proportional to ${r_{1}}^{7/3}$.

\subsubsection{Twisting motion}

The spring constant for the twisting displacement, $\phi$, is $k_{\rm t}$.
For the frictional model, we set
\begin{equation}
k_{\rm t} = \frac{16}{3} \mathcal{G}^{'} {a_{0}}^{3},
\end{equation}
where $\mathcal{G}^{'} = \mathcal{G} / 2$ is the reduced shear modulus.
We set $k_{\rm t} = 0$ for the frictionless model, as is the case for $k_{\rm r}$ and $k_{\rm s}$.
The critical angle for twisting, $\phi_{\rm crit}$, is set to $\phi_{\rm crit} = 1 / {( 16 \pi )}$ \citep{2007ApJ...661..320W}.
The energy needed to twist over $\pi / 2$ radian, $E_{\rm twist}$, is given by 
\begin{eqnarray}
E_{\rm twist} & = & k_{\rm t} \phi_{\rm crit} \frac{\pi}{2} \nonumber \\
              & = & \frac{1}{12} \mathcal{G} {a_{0}}^{3},
\end{eqnarray}
and $E_{\rm twist}$ is proportional to ${r_{1}}^{2}$.

\section{Collisional growth efficiency}
\label{app:1st}

\begin{figure*}
\centering
\includegraphics[width=0.48\textwidth]{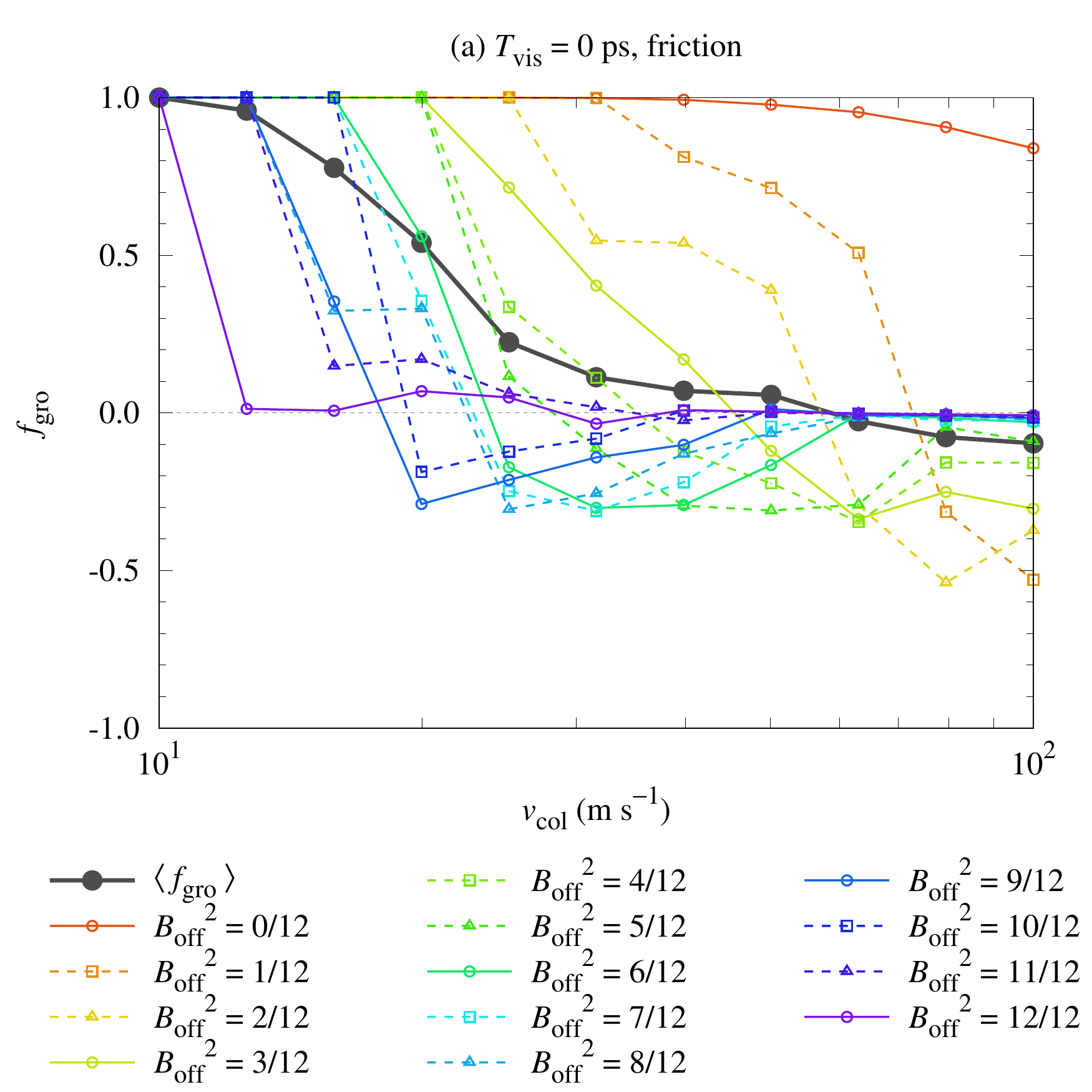}
\includegraphics[width=0.48\textwidth]{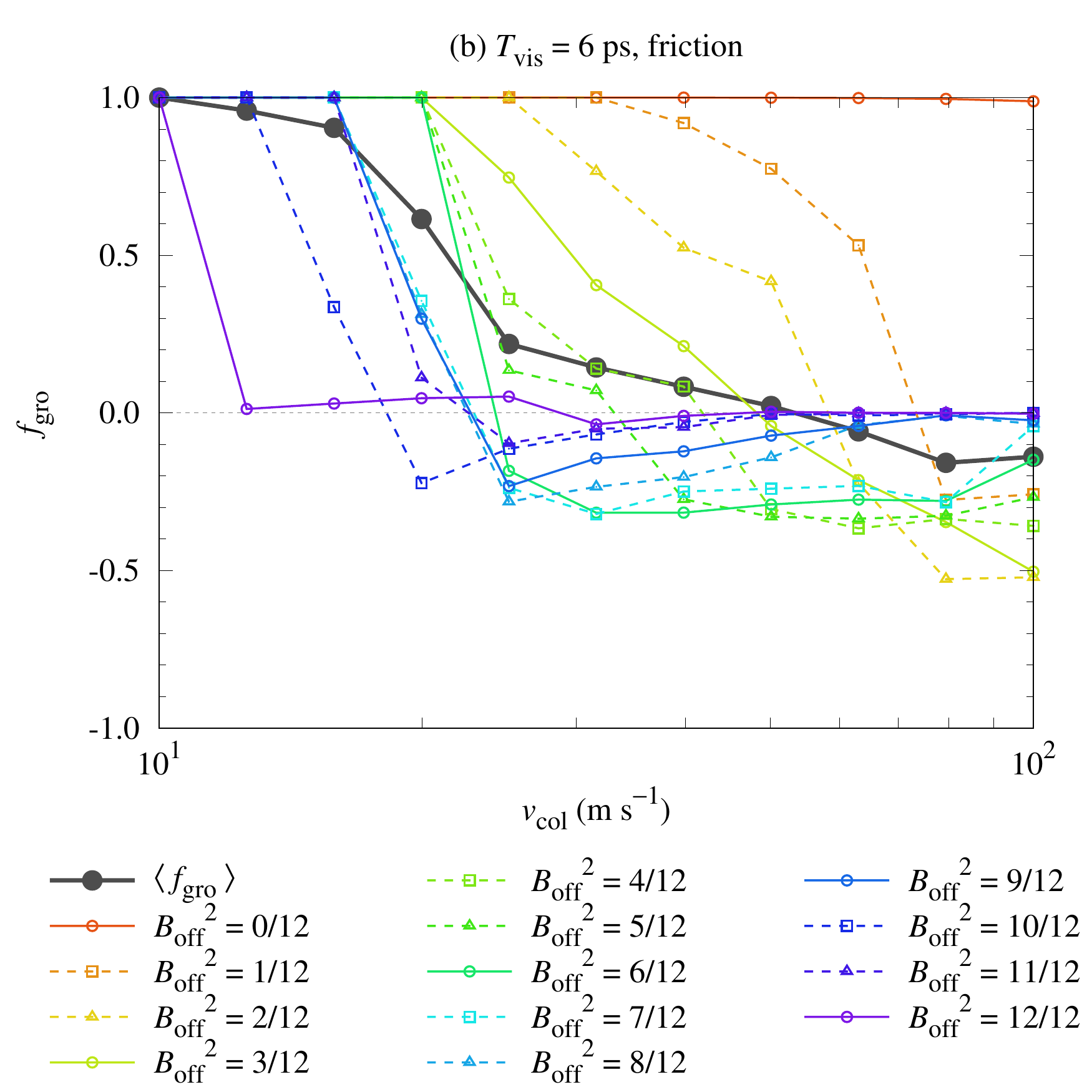}
\includegraphics[width=0.48\textwidth]{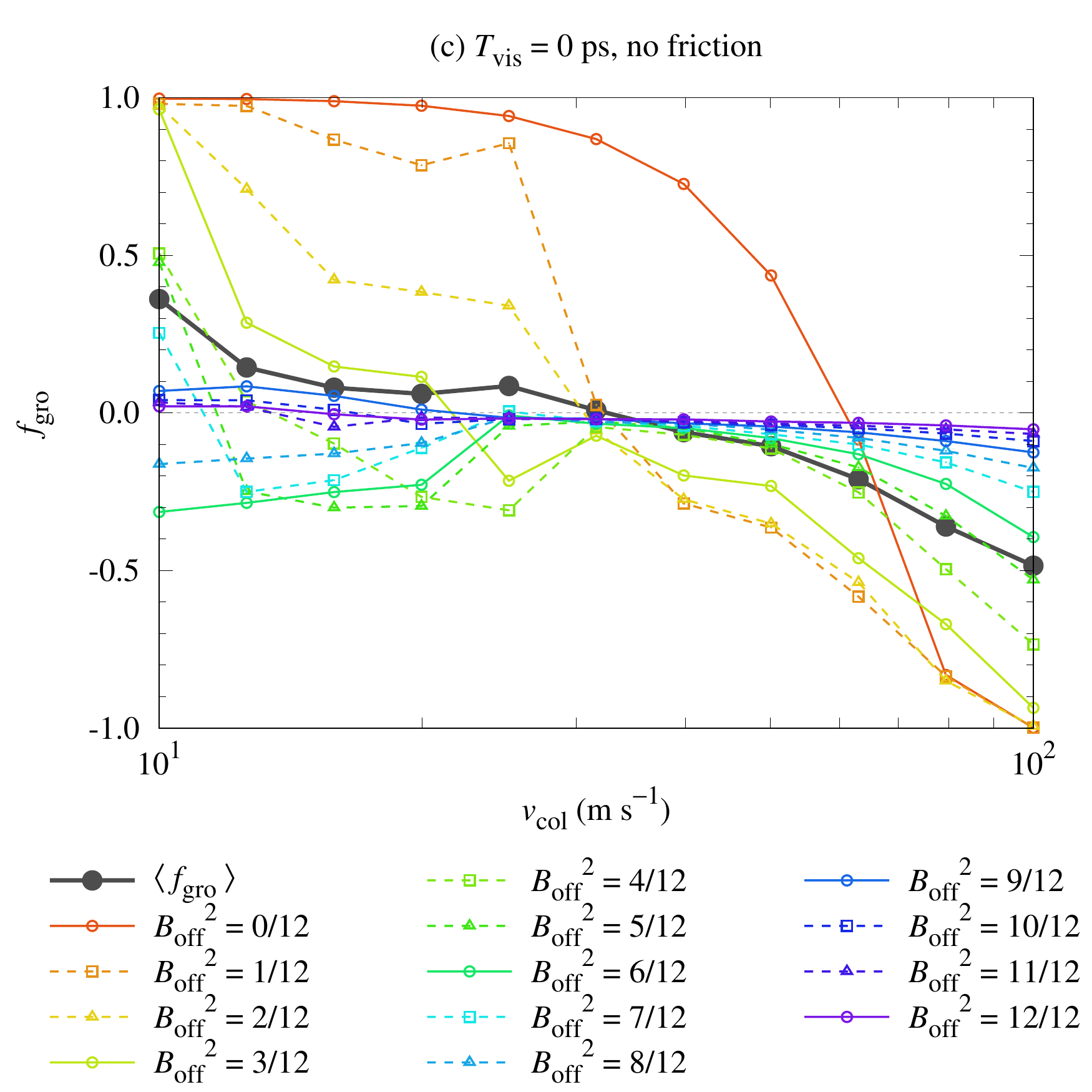}
\includegraphics[width=0.48\textwidth]{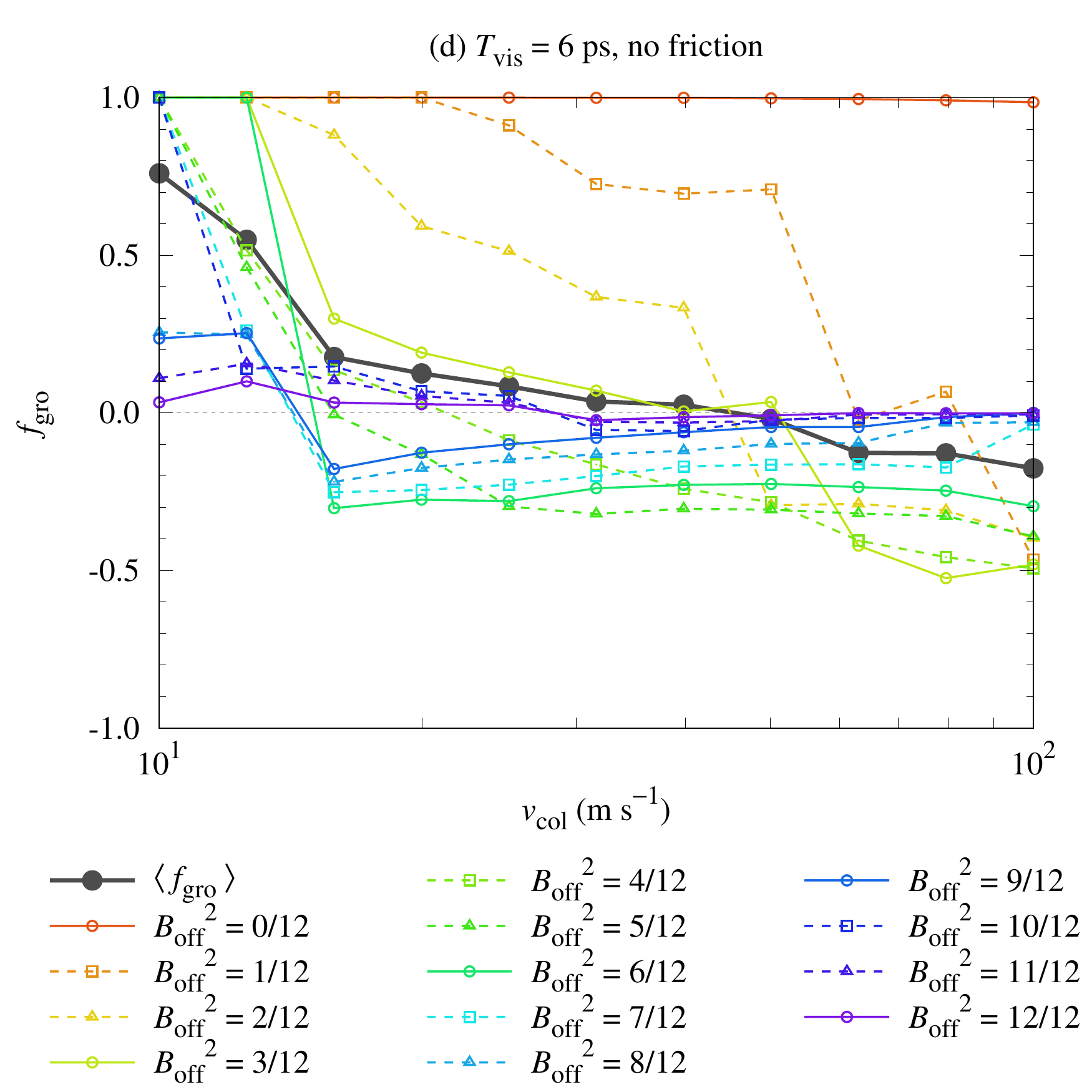}
\caption{
Collisional growth efficiency, $f_{\rm gro}$, for different settings of $B_{\rm off}$ and $v_{\rm col}$.
(a) For the frictional model without normal dissipation.
(b) For the frictional model with normal dissipation.
(c) For the frictionless model without normal dissipation.
(d) For the frictionless model with normal dissipation.
}
\label{fig:1st}
\end{figure*}

Figure \ref{fig:1st} shows the collisional growth efficiency, $f_{\rm gro}$.
The square of the normalized impact parameter ranges from ${B_{\rm off}}^{2} = 0$ to $1$ with an interval of $1/12$.
The collision velocity is set to $10^{( 0.1 i )}~\si{m}~\si{s}^{-1}$, where $i = 10$, $11$, ..., $20$.
Each panel shows $f_{\rm gro}$ for different particle interaction models.
The gray lines in Figure \ref{fig:1st} are the $B_{\rm off}$-weighted collisional growth efficiency, ${\langle f_{\rm gro} \rangle}$, and they are identical to those shown in Figure \ref{fig:1st_ave}.

\section{Energy dissipation and interparticle connection/disconnection}
\label{app:edis}

Here we briefly check the energy dissipation in our simulations (Appendix \ref{app:energy}).
In addition, we also show the numbers of connection and disconnection events (Appendix \ref{app:connection}).

\subsection{Energy dissipation}
\label{app:energy}

\begin{figure*}
\centering
\includegraphics[width=0.48\textwidth]{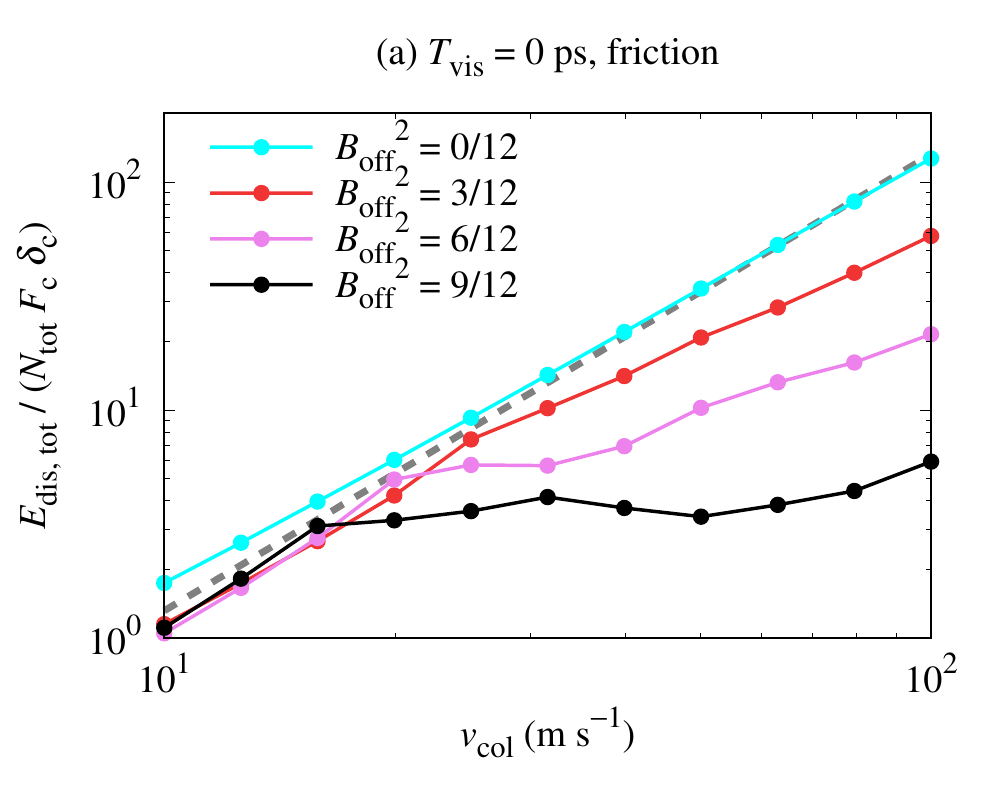}
\includegraphics[width=0.48\textwidth]{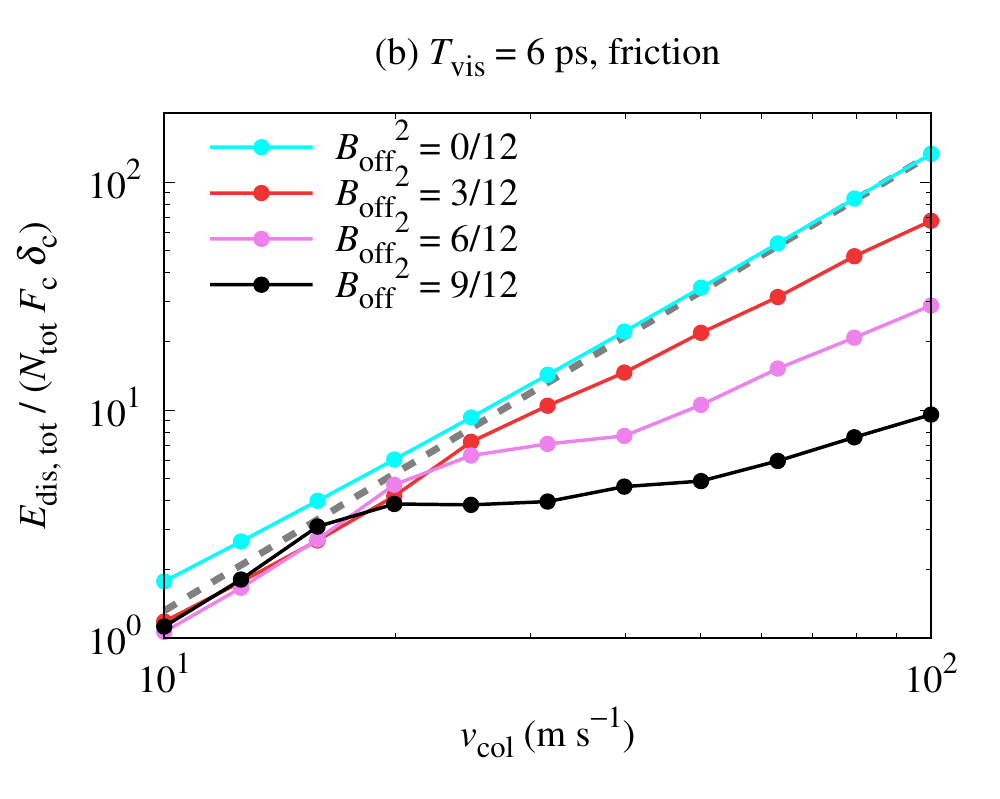}
\includegraphics[width=0.48\textwidth]{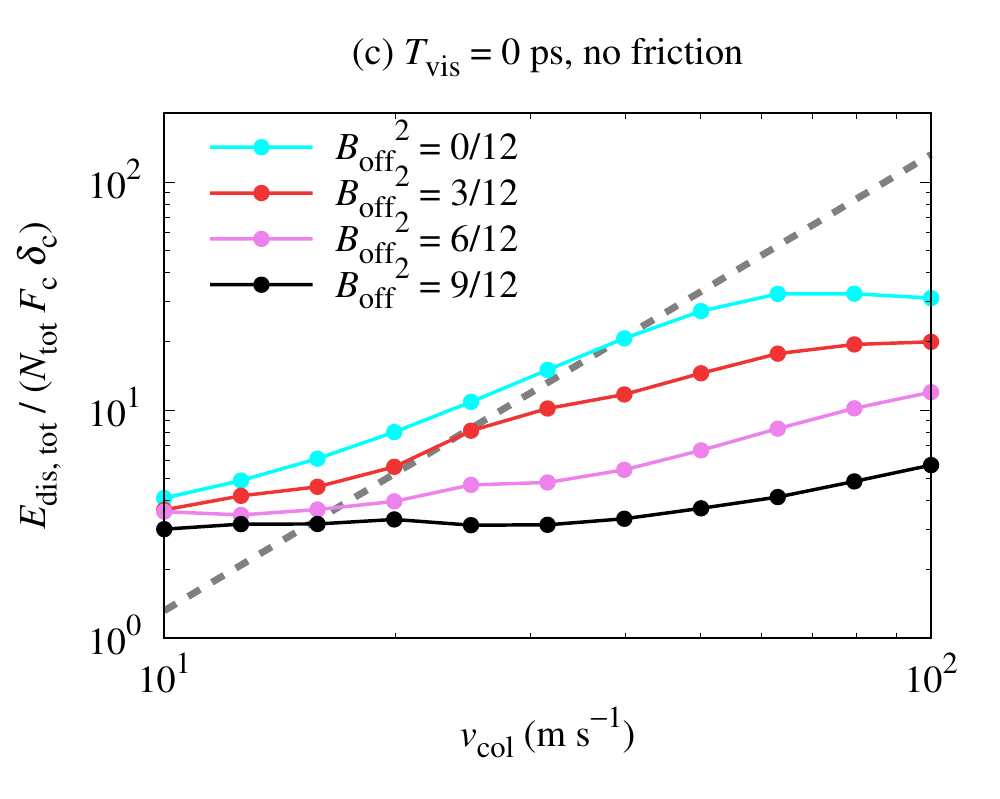}
\includegraphics[width=0.48\textwidth]{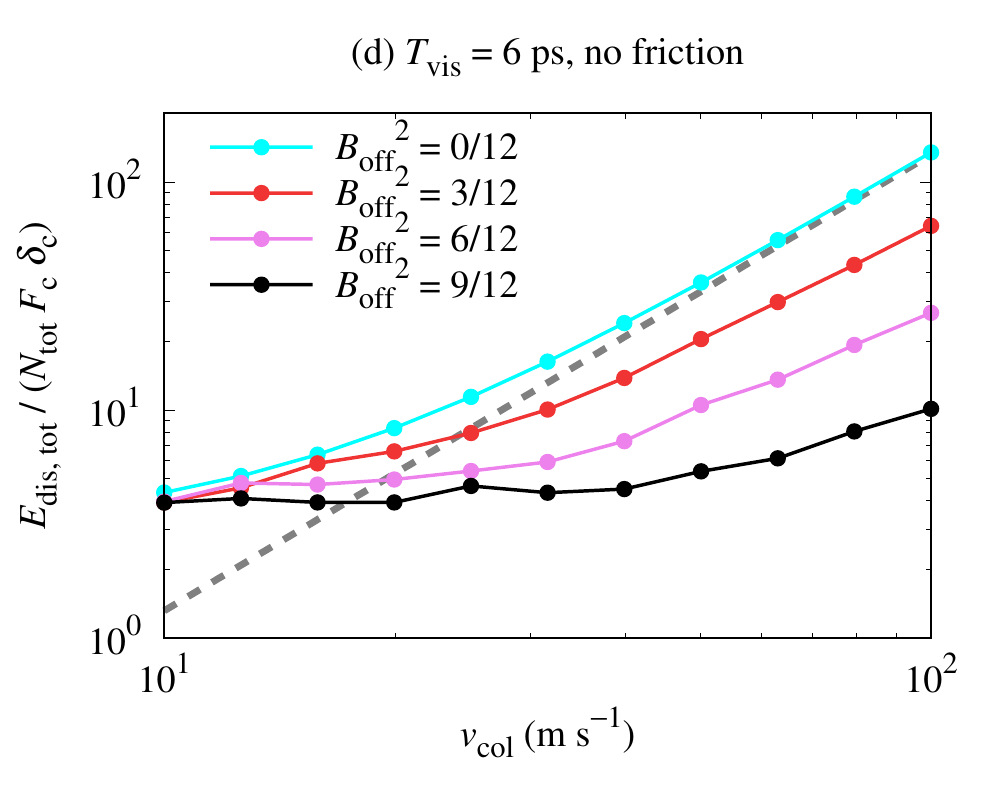}
\caption{
Total energy dissipation due to particle interactions from the start to the end of the simulations, $E_{\rm dis, tot}$, as a function of $v_{\rm rel}$ and $B_{\rm off}$.
(a) For the frictional model without normal dissipation.
(b) For the frictional model with normal dissipation.
(c) For the frictionless model without normal dissipation.
(d) For the frictionless model with normal dissipation.
The gray dashed line denotes the initial kinetic energy \citep[see][]{2022ApJ...939..100A}.
}
\label{fig:edis}
\end{figure*}

Figure \ref{fig:edis} shows the total energy dissipation due to particle interactions from the start to the end of the simulations, $E_{\rm dis, tot}$ \citep[see also Figure 8 of][]{2022ApJ...939..100A}.
For frictionless models, $E_{\rm dis, tot}$ is given by $E_{\rm dis, tot} = E_{\rm dis, c} + E_{\rm dis, v}$, where $E_{\rm dis, c}$ is the energy dissipation due to the connection and disconnection of particles and $E_{\rm dis, v}$ is the energy dissipation due to the viscous drag force.
The gray dashed line denotes the initial kinetic energy.

The relation between $E_{\rm dis, tot}$ and $f_{\rm gro}$ is complex.
At approximately $v_{\rm col} = 10~\si{m}~\si{s}^{-1}$, $E_{\rm dis, tot}$ for frictionless models is larger than that for frictional models; however, ${\langle f_{\rm gro} \rangle}$ for frictionless models is lower than that for frictional models (see Figure \ref{fig:1st_ave}).
For ${B_{\rm off}}^{2} = 0$ and $v_{\rm col} = 100~\si{m}~\si{s}^{-1}$, $E_{\rm dis, tot}$ is nearly equal to the initial kinetic energy in Figures \ref{fig:edis}(a), \ref{fig:edis}(b), and \ref{fig:edis}(d), and $f_{\rm gro} \sim 1$ in these cases (see Figure \ref{fig:1st}).
In contrast, $E_{\rm dis, tot}$ is significantly smaller than the initial kinetic energy in Figure \ref{fig:edis}(c), and the corresponding growth efficiency is $f_{\rm gro} \sim - 1$ in this case.

\begin{figure*}
\centering
\includegraphics[width=0.48\textwidth]{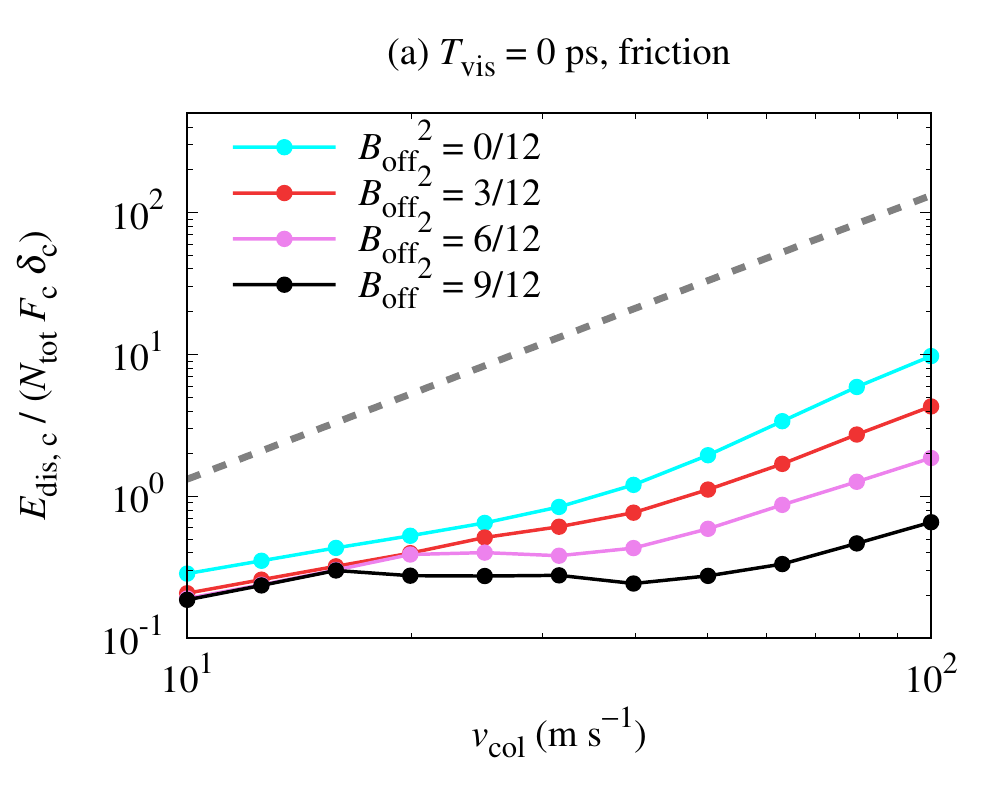}
\includegraphics[width=0.48\textwidth]{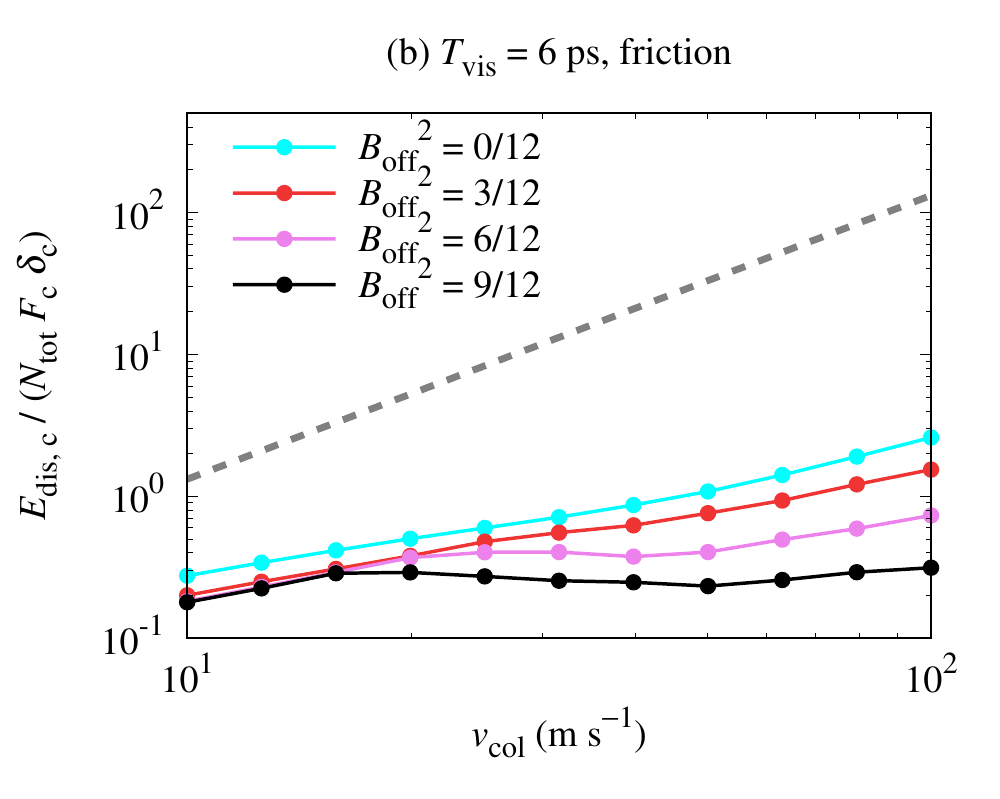}
\includegraphics[width=0.48\textwidth]{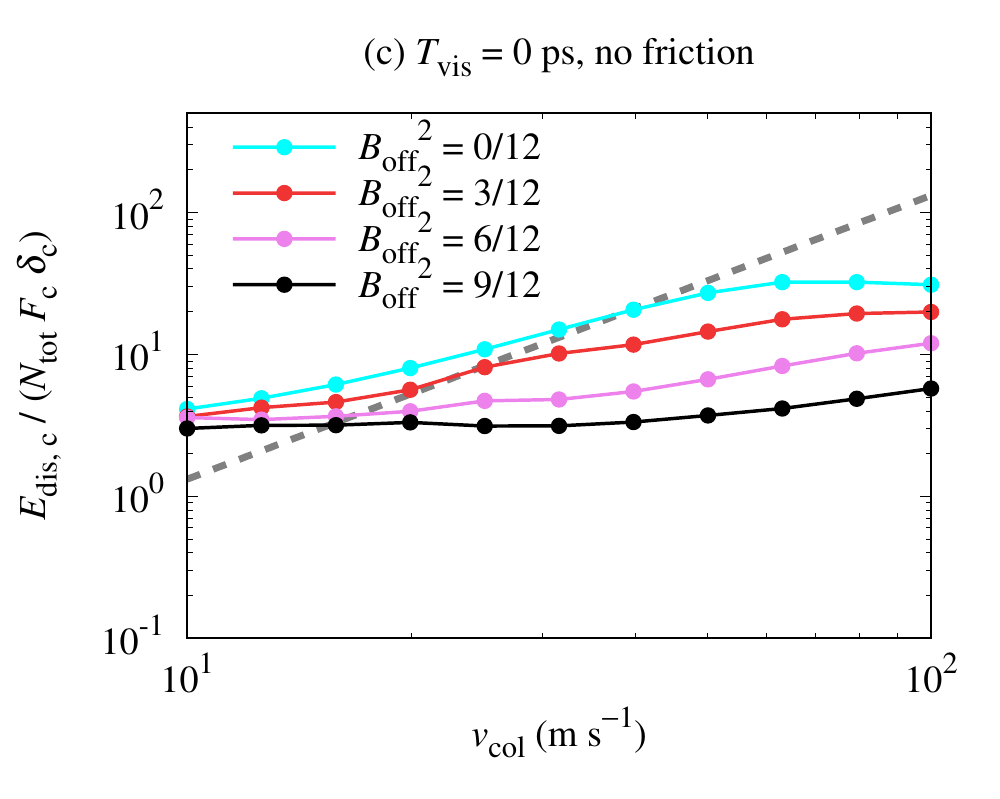}
\includegraphics[width=0.48\textwidth]{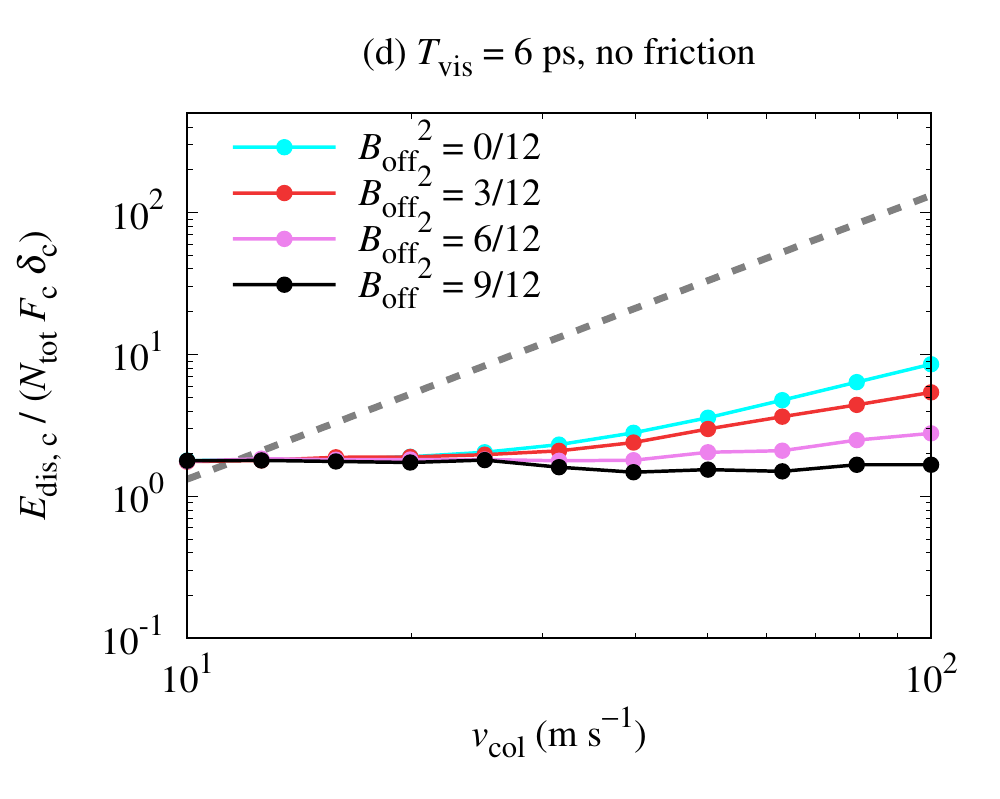}
\caption{
Energy dissipation due to connection and disconnection of particles, $E_{\rm dis, c}$, as a function of $v_{\rm rel}$ and $B_{\rm off}$.
(a) For the frictional model without normal dissipation.
(b) For the frictional model with normal dissipation.
(c) For the frictionless model without normal dissipation.
(d) For the frictionless model with normal dissipation.
The gray dashed line denotes the initial kinetic energy.
}
\label{fig:edisc}
\end{figure*}

Figure \ref{fig:edisc} shows $E_{\rm dis, c}$ as a function of $v_{\rm rel}$ and $B_{\rm off}$.
We note that $E_{\rm dis, c}$ is identical to $E_{\rm dis, tot}$ for the frictionless model without normal dissipation (Figure \ref{fig:edisc}(c)).
It is clear that $E_{\rm dis, c}$ for frictionless models is larger than that for frictional models in the range of $10~\si{m}~\si{s}^{-1} \le v_{\rm col} \le 100~\si{m}~\si{s}^{-1}$.
For frictional models, the fraction of energy dissipation due to the connection and disconnection of particles is notably small: $E_{\rm dis, c} / E_{\rm dis, tot} \ll 1$.
This is because the main energy dissipation mechanism for collisions of dust aggregates is the tangential friction between particles in contact for frictional models \citep{2022ApJ...933..144A, 2022ApJ...939..100A}.

\subsection{Interparticle connection and disconnection}
\label{app:connection}

\begin{figure*}
\centering
\includegraphics[width=0.48\textwidth]{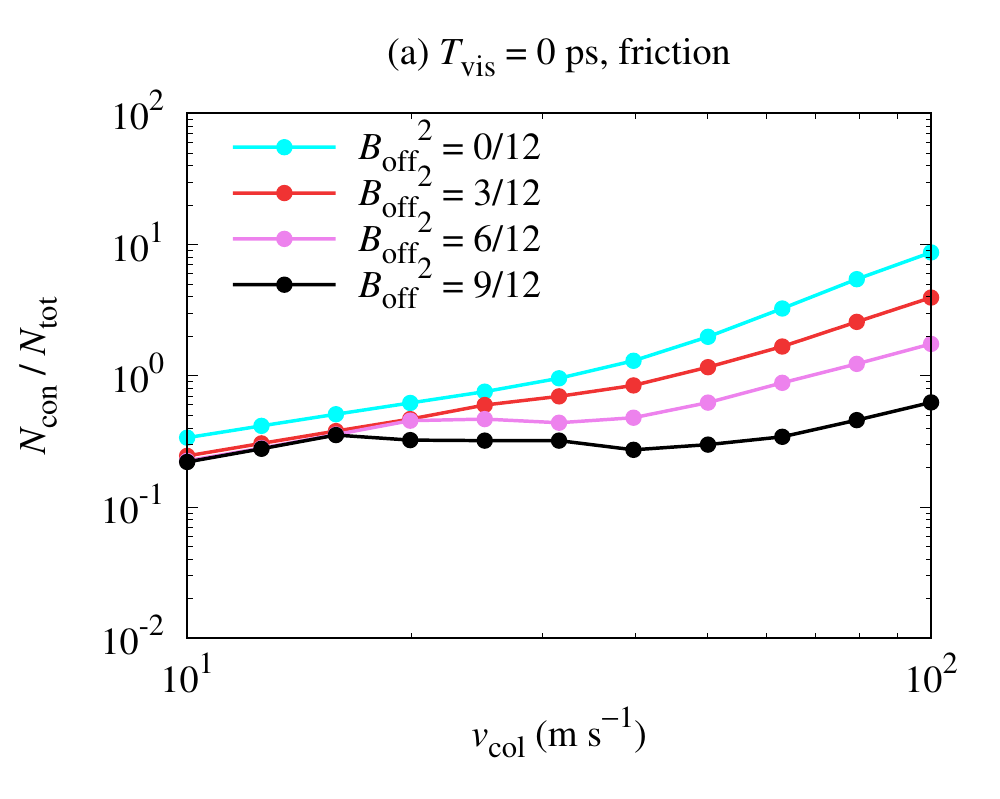}
\includegraphics[width=0.48\textwidth]{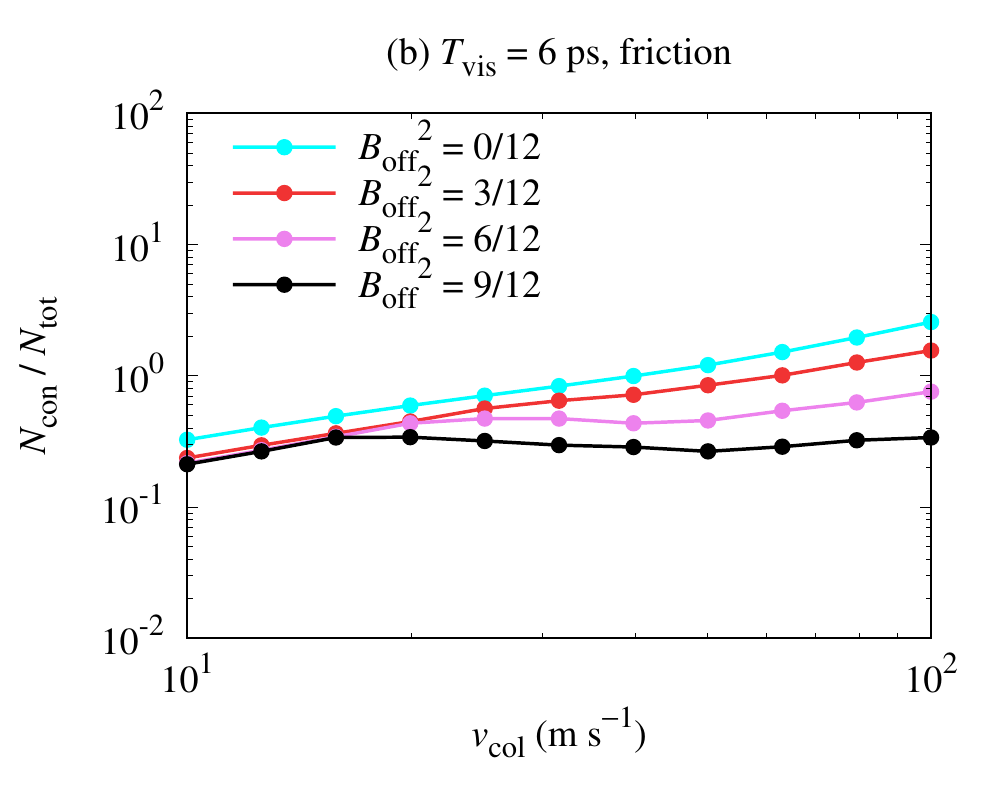}
\includegraphics[width=0.48\textwidth]{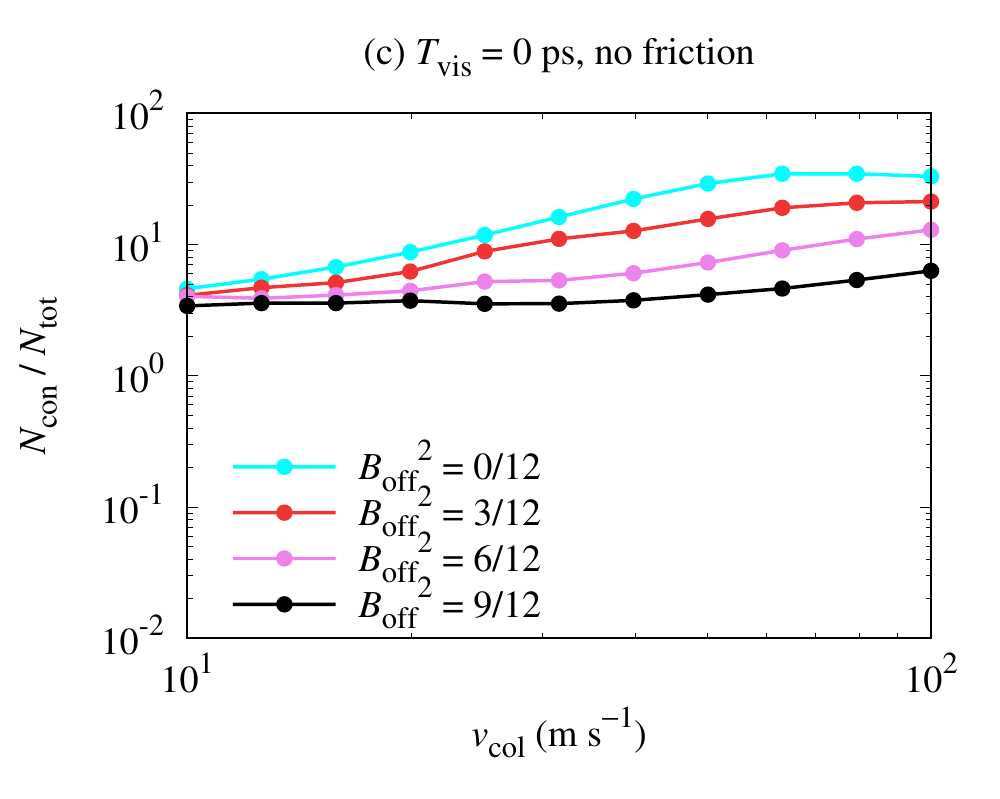}
\includegraphics[width=0.48\textwidth]{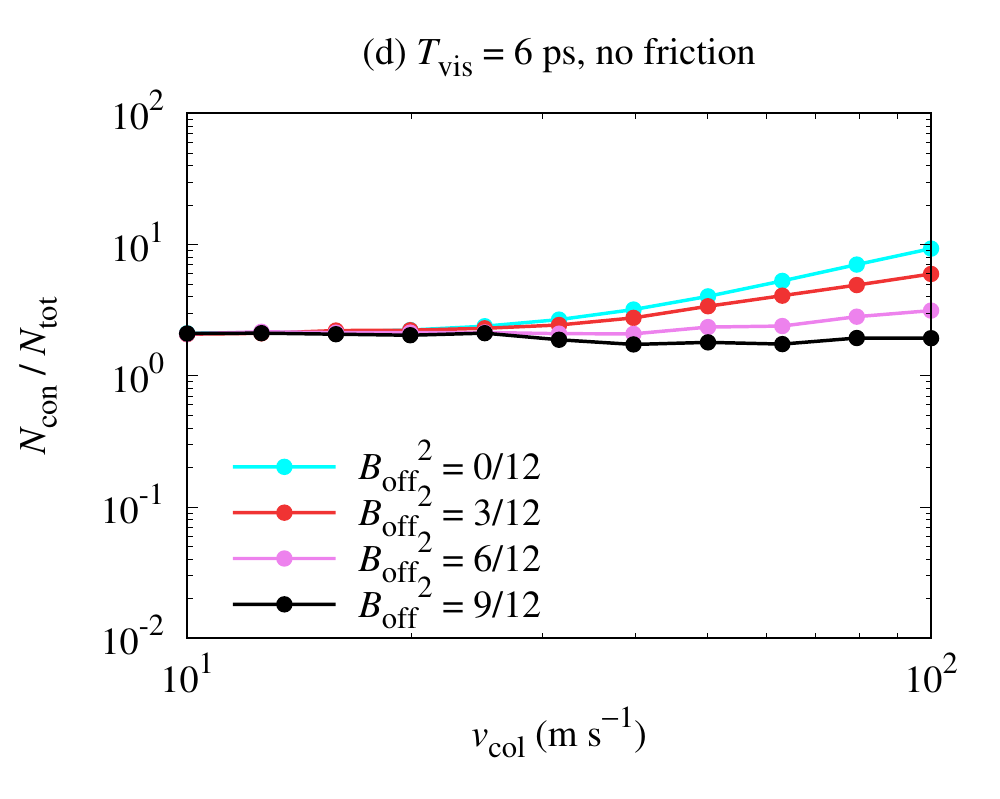}
\caption{
Number of connection events in a collision between dust aggregates, $N_{\rm con}$, as a function of $v_{\rm rel}$ and $B_{\rm off}$.
(a) For the frictional model without normal dissipation.
(b) For the frictional model with normal dissipation.
(c) For the frictionless model without normal dissipation.
(d) For the frictionless model with normal dissipation.
}
\label{fig:Ncon}
\end{figure*}

\begin{figure*}
\centering
\includegraphics[width=0.48\textwidth]{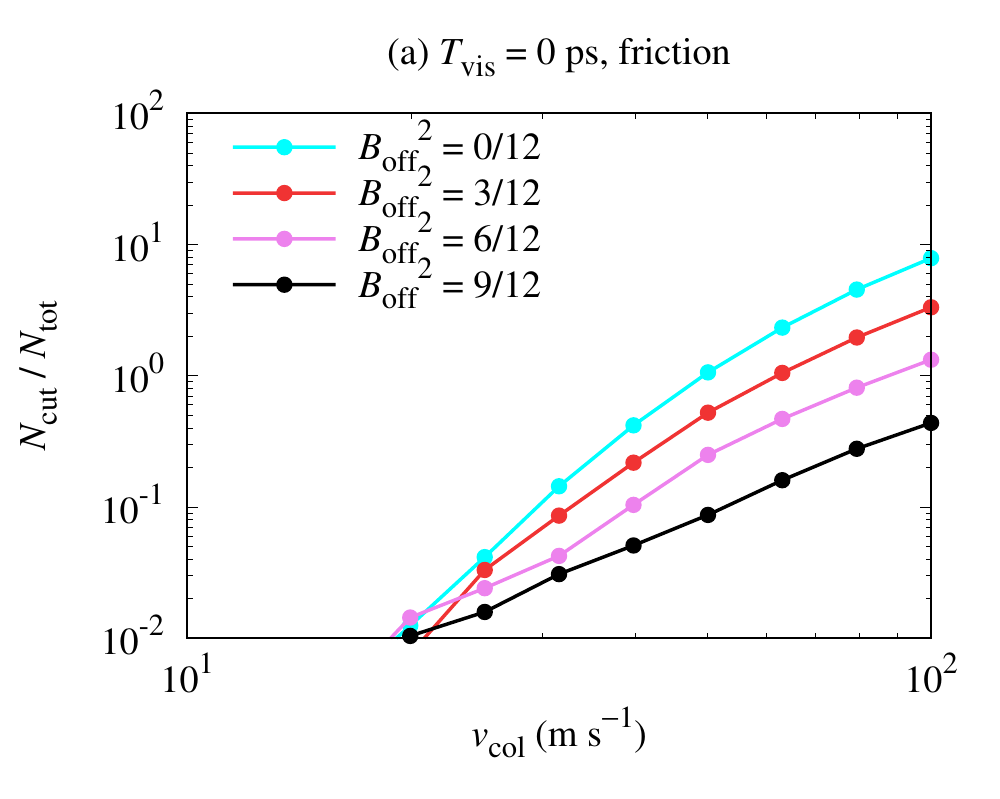}
\includegraphics[width=0.48\textwidth]{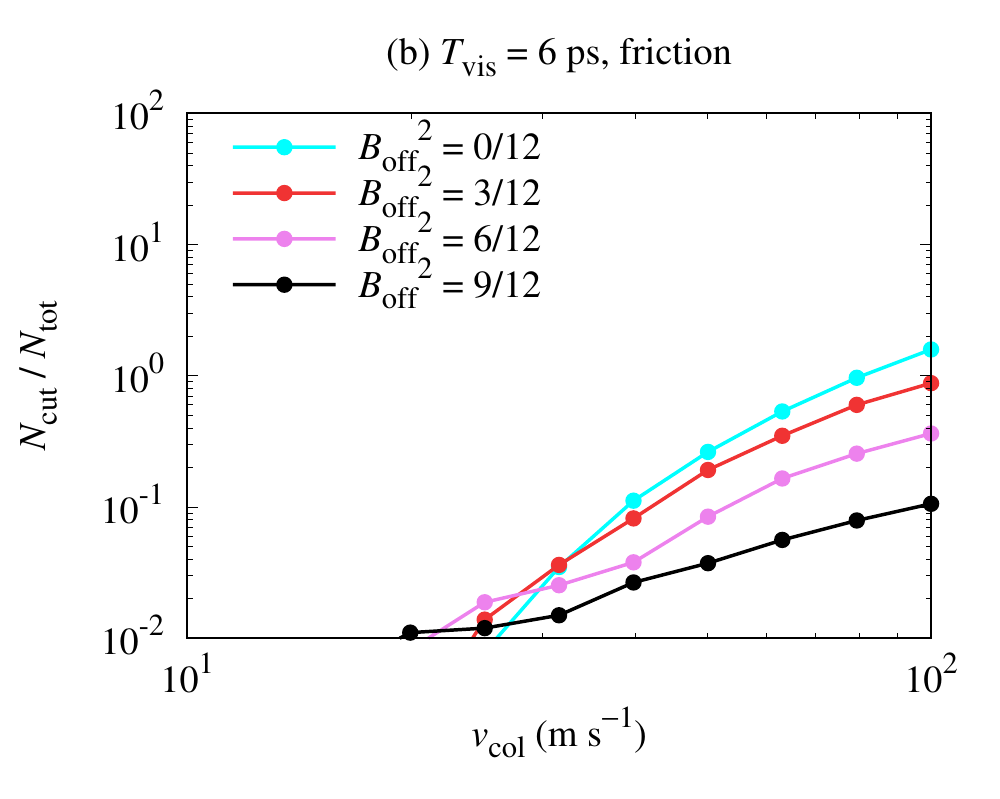}
\includegraphics[width=0.48\textwidth]{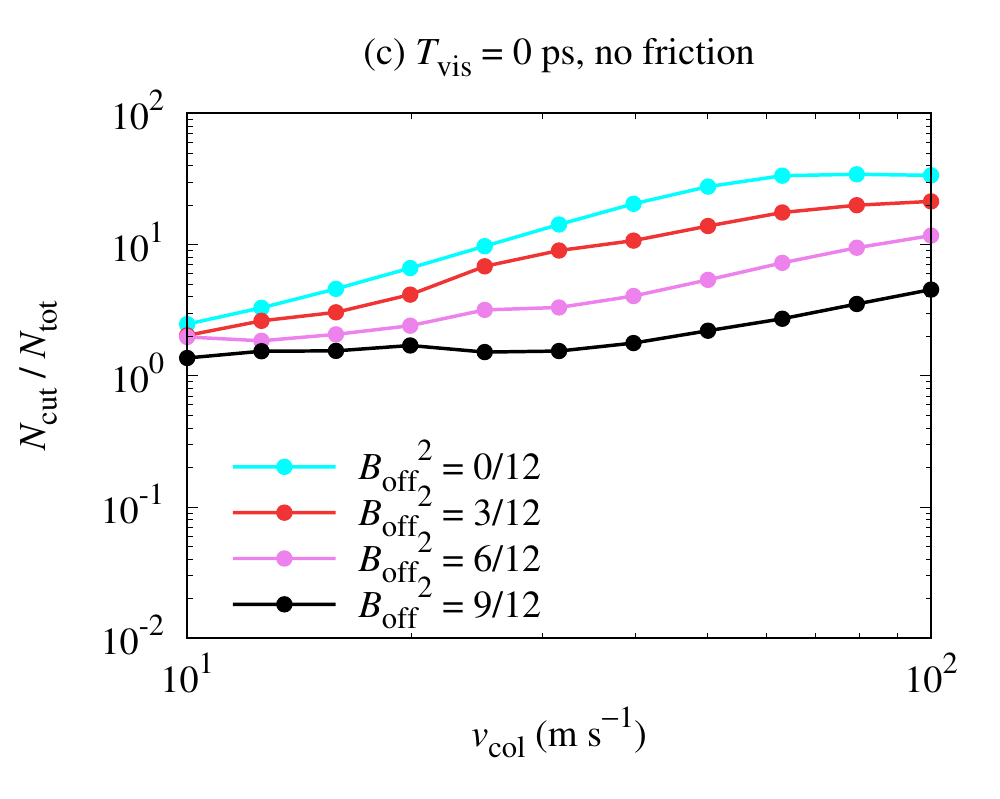}
\includegraphics[width=0.48\textwidth]{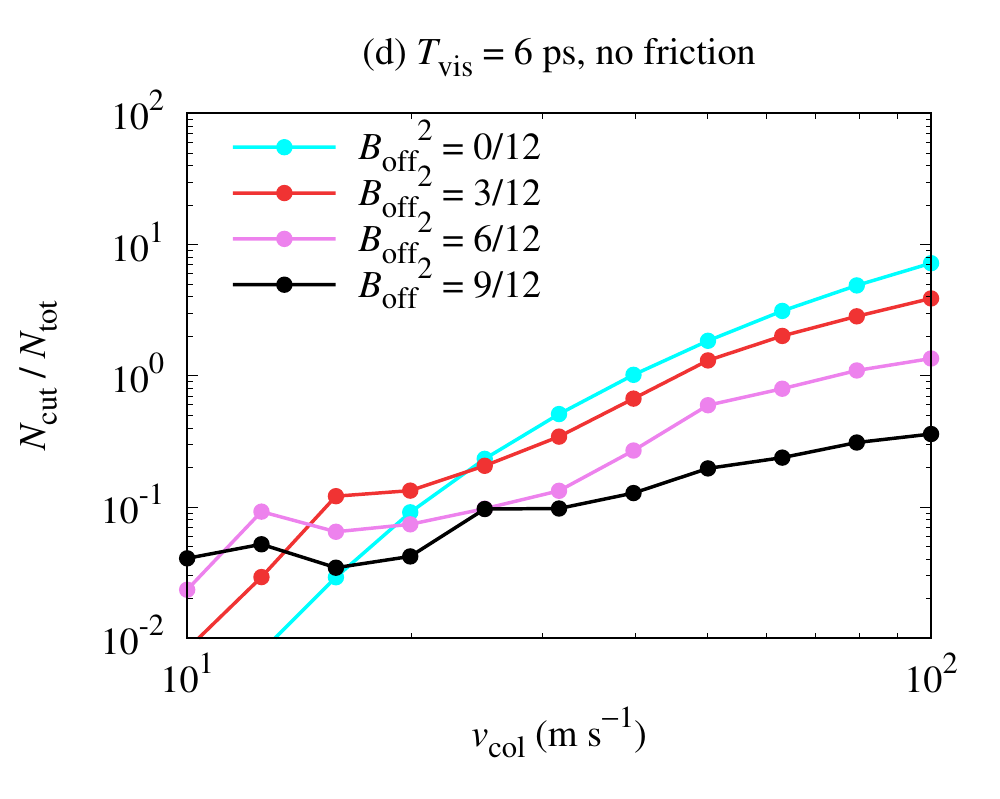}
\caption{
Number of disconnection events in a collision between dust aggregates, $N_{\rm cut}$, as a function of $v_{\rm rel}$ and $B_{\rm off}$.
(a) For the frictional model without normal dissipation.
(b) For the frictional model with normal dissipation.
(c) For the frictionless model without normal dissipation.
(d) For the frictionless model with normal dissipation.
}
\label{fig:Ncut}
\end{figure*}

We also check the numbers of connection and disconnection events, as they are directly related to $E_{\rm dis, c}$.
Figures \ref{fig:Ncon} and \ref{fig:Ncut} show the numbers of connection and disconnection events in a collision between dust aggregates, $N_{\rm con}$ and $N_{\rm cut}$, respectively \citep[see also Figure 15 of][]{2022ApJ...933..144A}.

For frictionless models, $N_{\rm con}$ is larger than $N_{\rm tot}$ in the range of $10~\si{m}~\si{s}^{-1} \le v_{\rm col} \le 100~\si{m}~\si{s}^{-1}$ (Figures \ref{fig:Ncon}(c) and \ref{fig:Ncon}(d)).
In contrast, $N_{\rm con}$ is smaller than $N_{\rm tot}$ for frictionless models at approximately $v_{\rm col} = 10~\si{m}~\si{s}^{-1}$ (Figures \ref{fig:Ncon}(a) and \ref{fig:Ncon}(b)).
The large difference in $N_{\rm con}$ would be related to the difference in the degree of deformation of aggregates after collisions.

For the frictionless model without normal dissipation, not only $N_{\rm con}$ but also $N_{\rm cut}$ exceeds $N_{\rm tot}$ in the entire range of $10~\si{m}~\si{s}^{-1} \le v_{\rm col} \le 100~\si{m}~\si{s}^{-1}$ (Figure \ref{fig:Ncut}(c)).
For other models, $N_{\rm cut}$ is significantly smaller than $N_{\rm tot}$ at approximately $v_{\rm col} = 10~\si{m}~\si{s}^{-1}$ (Figures \ref{fig:Ncut}(a), \ref{fig:Ncut}(b), and \ref{fig:Ncut}(d)).
As shown in Figure \ref{fig:1st_ave}, ${\langle f_{\rm gro} \rangle}$ for the frictionless model without normal dissipation is notably lower than those for the other three models.
We therefore imagine that the dependence of ${\langle f_{\rm gro} \rangle}$ on $v_{\rm col}$ might be the key to understanding the large difference in $v_{\rm fra}$ among particle interaction models.

\end{document}